\documentclass[%
reprint,
superscriptaddress,
twocolumn,
 amsmath,amssymb,
pra,
]{revtex4-2}

\usepackage{graphicx}
\usepackage{dcolumn}
\usepackage{bm}
\usepackage{hyperref}

\usepackage{mathrsfs}
\usepackage{color}
\usepackage{soul}
\usepackage{braket}

\usepackage{lipsum}
\usepackage{xargs} 
\usepackage[colorinlistoftodos,prependcaption,obeyDraft]{todonotes}
\usepackage{setspace}

\usepackage{float}
\usepackage{booktabs}

\newcommandx{\TODO}[2][1=]{\todo[inline,#1]{\begin{spacing}{1.0}#2\end{spacing}}}
\newcommandx{\commentMG}[2][1=]{\todo[linecolor=Plum,backgroundcolor=Plum!25,bordercolor=Plum,#1]{\begin{spacing}{1.0}MG:~#2\end{spacing}}}
\newcommandx{\commentVM}[2][1=]{\todo[linecolor=blue,backgroundcolor=blue!25,bordercolor=blue,#1]{VM:~#2}}
\newcommandx{\commentSC}[2][1=]{\todo[linecolor=OliveGreen,backgroundcolor=OliveGreen!25,bordercolor=OliveGreen,#1]{SC:~#2}}

\newcommand{\Op}[1]{\ensuremath{\hat{#1}}}

\newcommand{\Abs}[1]{\vert#1\vert}

\newcommand{\Avg}[1]{\ensuremath{\langle #1 \rangle}}

\begin{document}

\title{Extreme Spin Squeezing via Optimized One-Axis Twisting and Rotations}

\def\ARL{DEVCOM Army Research Laboratory, Adelphi, MD 20783}
\def\MIT{Department of Physics, MIT-Harvard Center for Ultracold Atoms, and Research Laboratory of Electronics, Massachusetts Institute of Technology, Cambridge, MA 02139}

\author{Sebastian C. Carrasco}
  \email{seba.carrasco.m@gmail.com}
  \affiliation{\ARL}

\author{Michael H. Goerz}
  \affiliation{\ARL}

\author{Zeyang Li}
  \affiliation{\MIT}

\author{Simone Colombo}
  \affiliation{\MIT}

\author{Vladan Vuleti\'c}
  \affiliation{\MIT}

\author{Vladimir S. Malinovsky}
  \affiliation{\ARL}

\date{\today}

\begin{abstract}
  We propose a novel scheme for the generation of optimal squeezed states for Ramsey interferometry. The scheme consists of an alternating series of one-axis twisting pulses and rotations, both of which are straightforward to implement experimentally. The resulting states show a metrological gain proportional to the Heisenberg limit. We demonstrate  that the Heisenberg scaling is maintained even when placing constraints on the amplitude of the pulses implementing the one-axis twisting and when taking into account realistic losses due to photon scattering.
\end{abstract}

\maketitle

\section{Introduction}

In quantum metrology, an ensemble of $N$ non-interacting atoms provides a basic statistical scaling of the measurement precision with $1/\sqrt{N}$, also known as the standard quantum limit (SQL). However, quantum mechanics also provides for the possibility of entanglement between the atoms. The non-classical correlations can provide an advantage in the scaling of the measurement precision, up to the Heisenberg limit (HL) with a scaling of $1/N$, that is, a quadratic advantage~\cite{degen2017quantum}. Overcoming the SQL is central in the effort of advanced quantum metrology that actively exploits the fundamental quantum properties of the measurement device. Approaching the Heisenberg limit promises dramatic improvements in high-precision sensing with a relatively small amount of atoms. Such high-precision sensing is the ultimate objective of applications that include the search for dark matter~\cite{derevianko2014hunting}, tests of the fundamental laws of physics~\cite{safronova2018search, safronova2019search}, gravitational waves detection~\cite{kolkowitz2016gravitational}, timekeeping~\cite{GuinotM2005}, and geodesy~\cite{mehlstaubler2018atomic, grotti2018geodesy, takamoto2020test}. Consequently, substantial effort has been devoted to design protocols that allow to achieve the limit.

A well-known strategy to reach a scaling beyond the SQL is the realization of spin squeezed states (SSS)~\cite{kitagawa1993squeezed, wineland1992spin, wineland1994squeezed, MaPR2011, QinN2020}. These states are characterized by non-classical correlations that lower (squeeze) the variance of one measurement quadrature in the collective state by increasing the variance of the quadrature orthogonal to the measurement. The earliest proposal to generate SSS is based on engineering an effective atom-atom interaction, described as a one-axis twisting (OAT) Hamiltonian~\cite{kitagawa1993squeezed}. The spin-squeezed states that can be realized via one-axis twisting achieve a scaling beyond the SQL, but fall short of achieving the full Heisenberg limit. This is even more true when taking into account realistic losses in a typical experimental setup~\cite{braverman2019near}.

A relevant figure of merit for a specific spin-squeezed state is the metrological gain, i.e., the attainable phase sensitivity $\Delta \varphi$ for the measurement of the accumulated phase $\varphi$ in a Ramsey interferometric measurement~\cite{berman1997atom, cronin2009optics} relative to $\Delta \varphi_{\text{CSS}} = 1 / \sqrt{N}$. We consider here spin-squeezed states that redistribute the variance in two orthogonal directions of the total angular momentum (an effective collective spin representing $N$ atoms). Maximizing metrological gain for the measurement of a particular momentum component allows to find optimal states for Ramsey interferometry. These squeezed states result in the Heisenberg
scaling~\cite{bloch2004control, toth2014quantum}; they are also known as extreme spin squeezed states~\cite{sorensen2001entanglement}. In this letter, we design a protocol to generate such extreme spin-squeezed states.

Our proposed scheme relies only on experimentally demonstrated OAT interactions in an alternating sequence with standard spin rotations. Thus, the advantage of our approach is that it can be realized in current experimental setups that employ OAT, specifically optical lattice clock atom experiments~\cite{leroux2010implementation, leroux2012unitary, braverman2019near, colombo2021time, li2021collective}. We find that only two applications of the OAT Hamiltonian are sufficient to generate Heisenberg scaling. Moreover, we can show that our approach retains this scaling in a dissipative environment of light-mediated interactions in an optical cavity as a means to create the effective OAT Hamiltonian~\cite{leroux2012unitary}.

\section{Generating Extreme Spin Squeezing}

We consider an ensemble of $N$ identical two-level systems, or equivalently $N$ spin-$\frac{1}{2}$ particles with total spin~$S$. The collective spin $\vec{S}$ of the $N$ atom system is defined by its components in terms of the Pauli matrices as $\Op{S}_i = \frac{1}{2} \sum_{n = 1}^N \Op{\sigma}_{i, n}$, where $\Op{\sigma}_{i, n}$ is the $i$'th Pauli matrix acting on the particle $n$ with $i = x, y, z$.
Introducing an effective interaction between atoms allows to generate correlations in the atomic ensemble and to prepare entangled states. In a Ramsey spectroscopic measurement, these entangled states can result in a higher sensitivity compared to uncorrelated atomic states.

\begin{figure}
    \centering
    \includegraphics{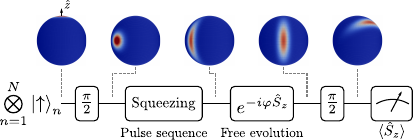}
    \caption{%
      \label{fig:Extreme_V2}
      Schematic of
the modified Ramsey interferometry scheme using spin squeezing. The spheres show the Husimi-Q function $\mathcal{Q}(\theta, \phi) = \Abs{\braket{\Psi | \theta, \phi }}^2$ over the generalized Bloch sphere, where $\ket{\theta, \phi}$ corresponds to a CSS oriented to the direction $(\theta, \phi)$ and $\ket{\Psi}$ is the state after each step in the scheme.
    }
\end{figure}
Figure~\ref{fig:Extreme_V2} 
illustrates the pulse sequence for a general Ramsey interferometer and the corresponding dynamics on the generalized Bloch sphere. The system is initialized to a coherent spin state (CSS). The CSS is usually created by optical pumping of all atoms into the ground state. A $\pi/2$-pulse then rotates the collective spin around the $y$ axis to the equator. At this point, an appropriate squeezing Hamiltonian may be used to produce a squeezed state with unequal quadratures of the distribution. Subsequently, the system evolves freely under the Hamiltonian $\delta \Op{S}_z$. This corresponds to a rotation around the $z$-axis of the generalized Bloch sphere, and accumulates a phase $\varphi$ directly proportional to the detuning $\delta$. A final $\pi/2$-pulse maps the phase $\varphi$ into a population difference between spin-up and spin-down states as measured by $\Avg{\Op{S}_z}$. The measurement allows to estimate $\varphi$ with sensitivity
\begin{equation}
  \label{eq:sensitivity}
  \Delta \varphi = \left| \frac{\Delta \Op{S}_z}{\partial \Avg{\Op{S}_z} / \partial \varphi} \right| \, ,
\end{equation}
evaluated for the final state.
Thus, squeezing of $\Delta \Op{S}_z$ in favor of a wider distribution along the $x$ and $y$ axes, as in the final state in Fig.~\ref{fig:Extreme_V2}, improves the precision of the phase estimation. In addition, we also have to consider the \emph{signal contrast}, which depends on the range of values for $\Avg{\Op{S}_z}$, with  $|\Avg{\Op{S}_z}| \le |\Avg{\vec S}|$. Since squeezing also lowers $|\Avg{\vec S}|$~\cite{MaPR2011}, phase sensitivity and signal contrast are generally not independent.

The metrological gain is given by the Wineland squeezing parameter~\cite{wineland1992spin, wineland1994squeezed}
\begin{equation}
  \label{eq:wineland}
  \xi^2
  = \frac{\Delta \varphi^2}{\Delta \varphi_{\text{CSS}}^2}
  = \underset{\hat u \perp \Avg{\vec{S}}}{\min} (\Delta \Op{S}_{\hat u}^2) \, \frac{N}{\Abs{\Avg{\vec{S}}}^2} \,,
\end{equation}
which relates the phase sensitivity without squeezing to the phase sensitivity with squeezing in Fig.~\ref{fig:Extreme_V2}. On the right-hand-side, the minimum is taken over all directions perpendicular to the mean spin direction $\Avg{\vec{S}}$.
For $\xi^2 = 1$, the measurement precision scales as the SQL with $1/\sqrt{N}$. For $\xi^2 < 1$, we have a squeezed state.
In particular, a sensitivity scaling of the form $\xi^2 = a N^{-1}$ would be proportional to the Heisenberg limit, and reach the exact Heisenberg limit for $a=1$.

We define an \emph{extreme spin-squeezed state} as the states that minimizes $\xi^2$, Eq.~\eqref{eq:wineland}, under the condition that the total wavefunction is normalized and that $\Abs{\Avg{\vec S}}$ is constant and thus the signal contrast is fixed. Consequently, these states allow to achieve the maximum sensitivity of the Ramsey interferometric measurements~\cite{sorensen2001entanglement, toth2014quantum}. To find these states, we apply variational calculus to minimize the functional
\begin{equation}
  \mathscr{L} [\ket{\Psi}] = N \frac{\Avg{\Op{S}_z^2}}{\Avg{\Op{S}_x}^2} - \lambda_1 \Braket{\Psi | \Psi} - \lambda_2 \Avg{\Op{S}_x} \, ,
\end{equation}
where $ \lambda_{1,2}$ are Lagrange multipliers and the average is taken over the state $\ket{\Psi}$.
Without loss of generality, we have chosen the mean spin direction to be $\Op{S}_x$, and the minimum perpendicular variance to be along the $z$ axis. Note that this is a different choice than in Fig.~\ref{fig:Extreme_V2}.

Using the Euler-Lagrange formalism, we obtain
\begin{equation}
  N \left[
    \frac{1}{\Avg{\Op{S}_x}^2} \Op{S}_z^2
    - 2 \frac{\Avg{\Op{S}_z^2}}{\Avg{\Op{S}_x}^3} \Op{S}_x
    - \frac{\lambda_2}{N} \Op{S}_x
  \right] \ket{\Psi}
  = \lambda_1 \ket{\Psi} \,,
\end{equation}
which is an eigenvalue equation that characterizes extreme spin-squeezed states. Specifically, we recognize extreme-spin-squeezed states as eigenfunctions of a Hamiltonian of the form
\begin{equation} \label{eq:eigenproblem}
  \Op{H} = \chi \Op{S}_z^2 - \Omega \Op{S}_x \,,
\end{equation}
where the ratio $\Omega / \chi$ is a function of $\Avg{\Op{S}_x}$. The extreme spin-squeezed state  $\ket{\Psi}$ has to be determined iteratively by solving for the $\Omega / \chi$ that gives the desired value of $\Avg{\Op{S}_x}$ when solving the eigenvalue equation with the Hamiltonian in Eq. \eqref{eq:eigenproblem}.

\begin{figure}
    \centering
    \includegraphics{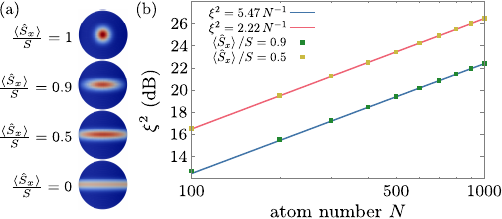}
    \caption{
      \label{fig:scaling}
      (a) Extreme spin-squeezed states with different mean spin lengths for $N=60$ atoms. The upper Bloch sphere shows the CSS for reference. 
      (b) Metrological gain for extreme spin-squeezed states with fixed mean spin length in terms of the Wineland squeezing parameter $\xi^2$, for varying number atoms $N$ (note the log-scale). We show the scaling for $\Avg{S_x} = 0.5 S$ and $\braket{S_x} = 0.9 S$, where $S$ is the total spin number, $S=N/2$. The solid lines are a linear fit showing the proportionality to the Heisenberg limit, $\xi^2 \propto N^{-1}$.
    }
\end{figure}

Figure~\ref{fig:scaling}~(a) shows the solution of the eigenvalue problem for different values of $\Avg{\Op{S}_x}$. As $\Avg{\Op{S}_x} \rightarrow 0$, the extreme spin-squeezed state converges to a ring on the equator of the Bloch sphere. In principle, this state provides the greatest metrological gain (minimal $\xi^2$). However, a very short spin length also implies a minimal signal in the experiment, which renders Ramsey interferometric measurements too susceptible to noise~\cite{toth2014quantum}.

It has been shown that all extreme spin squeezed states scale proportionally to the HL~\cite{toth2014quantum, rojo2003optimally}, and the constant of proportionality depends on the mean spin length, in our case on $\Avg{\Op{S}_x}$. We show examples of this scaling in Fig.~\ref{fig:scaling}~(b) for two values of $\Avg{\Op{S}_x}$.

The key to creating extreme spin-squeezed states is in the Hamiltonian Eq.~\eqref{eq:eigenproblem} of which they are eigenfunctions. Remarkably, up to one extra term, this Hamiltonian has the form of a one-axis twisting Hamiltonian~\cite{kitagawa1993squeezed} which is easily implemented experimentally~\cite{leroux2010implementation, leroux2012unitary, braverman2019near, colombo2021time}. The extra term is a rotation of the collective spin in a perpendicular direction.

Since the extreme spin-squeezed state is the ground state of the Hamiltonian in Eq.~\eqref{eq:eigenproblem}, one possibility for generating the state is by adiabatic time evolution. This process would start from a coherent spin state pointing along the $x$ axis~\cite{sorensen2001entanglement}. Considering the parameters $\chi$ and $\Omega$ as time-dependent control function would allow to slowly turn on the one-axis twisting, tuning the Hamiltonian from $\Op{H} = - \Omega \Op{S}_x$ to $\Op{H} = \chi \Op{S}_z^2 - \Omega \Op{S}_x$. Unfortunately, this process is too slow to be practical. A possible approach to overcome the slow adiabatic process is to implement the well-known ``shortcuts to adiabaticity''~(STA)~\cite{DelCampoPRL2013}, which would allow to quickly reach the target state, at least approximately~\cite{opatrny2016counterdiabatic, pichler2016noise}. However, an extra complication is that some experimental setups generate the OAT evolution by a series of pulses, which includes additional steps to cancel additional detrimental terms in the effective Hamiltonian. Incorporating these refocusing pulses in an STA approach would not be straightforward.

\begin{figure}
    \centering
    \includegraphics{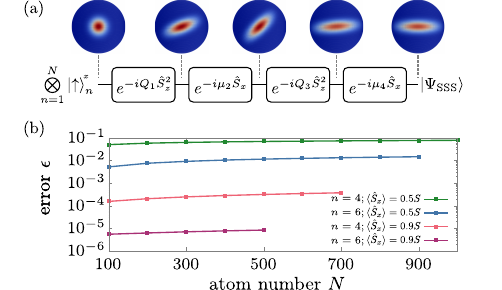}
    \caption{%
      \label{fig:sequence}
      Extreme spin squeezed states generation.
      (a) Snapshots of the extreme SSS after the each pulse in the squeezing sequence; $\Avg{\Op{S}_x}/S = 0.9$, $N=60$.
      (b) Infidelity of the generated extreme SSSs as a function of $N$; the mean spin length $\Avg{\Op{S}_x} = 0.5 S$ and $\Avg{\Op{S}_x} = 0.9 S$.
    }
\end{figure}
Here, we take an entirely different approach based on the realization that the two terms in the Hamiltonian in Eq.~\eqref{eq:eigenproblem} are individually easy to realize in an experiment.
The first term is the Hamiltonian that implements OAT, whereas the second term is a simple rotation in the extended Bloch sphere.
Hence, we propose a sequence of $n$ alternating OAT and $\Op{S}_x$ pulses that allow the creation of extreme spin-squeezed states without any fundamental alterations to existing experimental implementations for an OAT setup. Figure~\ref{fig:sequence}~(a) shows the proposed pulse sequence for an example of $N=60$ atoms initially in a CSS pointing along the $x$-axis. As shown, only $n=4$ pulses are sufficient to drive the CSS into an extreme spin-squeezed states with $\Avg{\Op{S}_x} = 0.9 S$. The extreme spin-squeezed state is reached with error $\epsilon=1 - |\eta|^2 \approx 1.3 \times 10^{-4}$, where $\eta=\braket{\Psi | \Psi_\text{SSS}}$ is the overlap between the final and target state $\ket{\Psi_\text{SSS}}$. To find the optimal pulse sequence, we propagate the initial CSS as
\begin{equation} \label{eq:seq}
\begin{split}
  \ket{\Psi_{\text{SSS}}}
  &=
  \exp \left(- i \mu_{n} \Op{S}_x \right)
  \exp \left(- i Q_{n-1} \Op{S}_z^2 \right)\,
  \dots
  \\
  & \quad
  \dots\,
  \exp \left(- i \mu_2 \Op{S}_x \right)
  \exp \left(- i Q_1 \Op{S}_z^2 \right)
  \ket{\Psi_\text{CSS}}\,,
\end{split}
\end{equation}
where $Q_k = \chi \, \Delta t$ is the shearing strength of $k$'th OAT pulse and $\mu_k=\Omega \Delta t$ is the rotation angle produced by the $k$'th $S_x$ pulse, $\Delta t$ is the pulse duration. We use SciPy's minimization routine~\cite{VirtanenNM2020} with the L-BFGS-B method~\cite{ZhuATMS97,ByrdSJSC1995} to iteratively find the parameters $(Q_k, \mu_k)$ that minimize $\epsilon$.

In Fig.~\ref{fig:sequence}~(b), we show the achieved error $\epsilon$ as a function of the number of atoms $N$ for different values of the contrast $\Avg{\Op{S}_x}/S = 0.5$ and $\Avg{\Op{S}_x}/S = 0.9$, using a sequence of four and six pulses, respectively. Unsurprisingly, higher contrast allows for higher fidelities. This is because the extreme spin-squeezed state is closer to the initial state. At the same time, the scaling of the error, respectively the loss of metrological gain due to infidelity, is nearly flat with respect to the number of atoms. Moreover, the fidelity also improves by increasing the number of pulses. Remarkably, we find that when generating the extreme spin-squeezed state with $\Avg{\Op{S}_x} = 0.9 S$ with four pulses, the optimization error introduces a metrological loss of less than $0.05$~dB compared to the result in Fig.~\ref{fig:scaling}~(b). The loss of precision due to the optimization error is higher for $\Avg{\Op{S}_x} = 0.5 S$ (between 3 and 6~dB). Note that the four-pulse sequence ($n = 4$) is the first non-trivial case, since the two-pulse sequence is equivalent to the OAT scheme, where the noise reduction scales proportionally to $N^{-2/3}$~\cite{kitagawa1993squeezed}.

\section{Metrological Loss due to Photon Scattering}

Optical methods to generate collective entanglement rely on the interaction of the atomic ensemble with a cavity light field. Cavity feedback squeezing~\cite{leroux2010implementation,li2021collective}  is a deterministic technique to generate a control Hamiltonian $\chi \Op{S}_z^2$. The spin quantum noise tunes the atom-cavity resonance such that the intracavity light intensity is proportional to the $z$ component of the collective atomic spin. Thus, the light induces an $\Op{S}_z$-dependent light-shift which generates $\Op{S}_y$-$\Op{S}_z$ quantum correlations and atomic entanglement in the process. Due to the cavity-enhanced atom-light interaction, any information contained in the light field results in the non-unitary evolution of the atomic system~\cite{liu2011spin}. Moreover, any photon scattered in the free space projects one atom into either spin-up or spin-down, thus reducing the spin coherence. However, as it has been recently demonstrated~\cite{colombo2021time,li2021collective} and modeled~\cite{liu2011spin,ZhangPRA2015},
decoherence can be minimized to reach a near-unitary evolution by tuning the entangling-light frequency. The sign of the $\chi \Op{S}_z^2$ Hamiltonian can be reversed simply by switching the sign of the detuning between the atom-cavity resonance and the light~\cite{li2021collective,liu2011spin}.

One way to produce an effective OAT is by the interaction of the atomic ensemble with a single-mode cavity field~\cite{schleier2010squeezing, leroux2010implementation, leroux2012unitary, braverman2019near, colombo2021time}. In this system,
\begin{equation}
  \Op{H} = - \hbar \omega n_c \left(\Op{S}_z + S\right) \, ,
\end{equation}
where $\omega$ represents the light shift per photon on the spin levels induced by the coupling to the cavity mode and $n_c$ is the number of photons~\cite{li2021collective}. If $\Avg{\Op{S}_z} = 0$ one can expand the photon number $\Avg{\hat n_c}(\Op{S}_z)$ in terms of the powers of $\Op{S}_z$ and approximate
\begin{align}
  \Op{H}
  &\approx
    - \hbar \omega \braket{n_c} (\Op{S}_z + S)
    - \hbar \omega \left<\frac{\partial \hat n_c}{\partial \Op{S}_z}\right>_{S_z = 0} \Op{S}_z^2
    \nonumber \\
  &=
    \alpha (\Op{S}_z + S) + \chi \Op{S}_z^2 \,.
    \label{Eq:Vladan}
\end{align}
Note that this is essentially the OAT Hamiltonian plus the $z$-rotation term $\alpha \Op{S}_z$; the $\alpha S$ term produces only a global phase and thus can be neglected.

The OAT Hamiltonian can clearly be implemented
by turning on the non-linear interaction shown in Eq.~(\ref{Eq:Vladan}) for a specific time, $\Delta t$, then applying a $\pi$-pulse, and finally turning on the non-linear interaction again. Effectively, the result of this sequence is identical to the application of the OAT Hamiltonian
%
  $\Op{H} = \chi \Op{S}_z^2$
%
(with $Q = \chi \Delta t$), 
since the $\pi$-pulse cancels the other terms, similar to a spin echo sequence. Applying $S_x$ rotation simultaneously with the OAT Hamiltonian may be challenging, as it affects the spin echo effect.
The optimized squeezing sequence of our proposal is not susceptible to this issue as the rotations are between each application of the OAT Hamiltonian, \emph{i.e.}, between each iteration of the described sequence.

As we stated above, during each application of the effective OAT, the photon scattering reduces the spin coherence, leading to a contrast loss at the final measurement. It was shown~\cite{colombo2021time} that the contrast loss, $C_\text{sc}$, depends on both the total shearing strength $Q=\sum_k Q_k$ of the pulse sequence (see Eq. \eqref{eq:seq}) and the number of atoms $N$. This can be parametrized as $C_\text{sc} = \exp ( - \gamma \tilde Q)$, where $\tilde Q = \sqrt{N} \, Q$ is the normalized shearing strength and $\gamma$ is the scaling parameter which we choose to be equal to $0.36$ addressing the experimental conditions reported in~\cite{colombo2021time}. This reduces the above calculated metrological gain to ${\tilde \xi}^2 = \xi^2 / C_\text{sc}^2$,
potentially affecting the precision scaling.

A primary advantage of our scheme is that it involves only minimal modifications to an experimental setup realizing OAT and uses only a small number of OAT applications, reducing the OAT-associated contrast loss.
This is in contrast to other approaches that have been proposed, e.g., the combination of OAT interactions and rotations for the creation of effective two-axis twisting~\cite{liu2011spin, shen2013efficient, zhang2014quantum, huang2015two, HuangNPJQI2021, HuPRA2017}, or modulating the OAT interaction and rotations to implement extreme-spin-squeezed states~\cite{sorensen2001entanglement, opatrny2012spin, zhang2014quantum, opatrny2016counterdiabatic}. Moreover, we can show that our scheme is robust in the presence of photon scattering.

Analyzing the extreme spin-squeezed states and the optimal pulse sequences that generate them, we observe that the states with lower $\Avg{\Op{S}_x}/\Op{S}$ provide higher metrological gain and creating them requires higher shearing strength. However, a reduction of the metrological gain due to the photon-scattering contrast loss is higher for the extreme SSSs with low $\Avg{\Op{S}_x}/S$. Therefore, as an example of a good trade-off between the level of squeezing and shearing strength
we consider the case $\Avg{\Op{S}_x} = 0.9 S$ in the following analysis.
In this case, applying only $n = 4$ pulses, the optimized extreme SSSs result in an error on the order of $10^{-4}$, see Fig.~\ref{fig:sequence}~(b).

\begin{figure}
    \centering
    \includegraphics{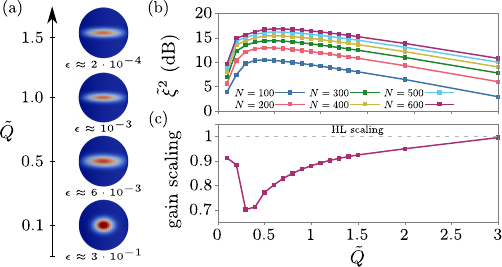}
    \caption{
      \label{fig:HL}
      Extreme SSS generation with fixed normalized shearing strength.
      (a) Approximated extreme SSSs generated by pulse sequence with fixed shearing strength, for $N = 100$ and $\Avg{\Op{S}_x} = 0.9 \Op{S}$.
      (b) Metrological gain, corrected for contrast loss due to photon scattering, as a function the normalized shearing strength, for different values of $N$.
      (c) Metrological gain scaling ($b$ from fit $\tilde \xi^2 = a  N^{-b}$) as a function the normalized shearing strength.
    }
\end{figure}
One way to minimize the effect of the contrast loss in the precision scaling is to fix $\tilde Q$ during the optimization of the pulse sequence. As a result, the contrast loss  $C_\text{sc}$ affects only the proportionality factor, but not the scaling. However, this reduces the metrological gain due to a reduction in the fidelity of the prepared state, so a scaling proportional to the Heisenberg limit is achievable only for sufficiently high values of $\tilde Q$. Fortunately, the convergence to the extreme spin-squeezed state is fast with an increase of $\tilde Q$, as we show in Fig.~\ref{fig:HL}~(a). The necessary shearing strength is of the order of one, $\tilde Q \sim 1$, as demonstrated experimentally in previous work~\cite{colombo2021time}. In Fig.~\ref{fig:HL}~(b), we plot the metrological gain corrected by the contrast loss due to photon scattering as a function of $\tilde Q$ for different values of $N$. We observe a peak around $\sqrt{N} \, Q = 0.5$ where the trade-off between contrast loss and precision is optimal. The position of the optimal precision is moving to larger values of shearing strength $\tilde Q$ as we increase the number of atoms, $N$.
We obtain 14.9 dB of metrological gain for $N = 350$ atoms using $\tilde Q = 0.55$,  which surpasses the result reported in~Ref.~\cite{colombo2021time}. Finally, in Fig.~\ref{fig:HL}~(c), we show the scaling of the metrological gain as a function of $\tilde Q$. As expected, it approaches the Heisenberg limit as we increase $\tilde Q$. Thus, the proposed optimized squeezing sequence achieves a scaling proportional to the HL.

We further analyze the robustness of the the scheme with respect to variations in the amplitude of the control pulses. To this end, we have applied random variations on the amplitude of both the OAT pulses and the rotations, drawn from a Gaussian distribution with a standard deviation of 10\% of the optimal amplitude. We find that the resulting standard deviation of the resulting metrological gain is less than 0.2~dB. Thus, our scheme is remarkably robust with respect to variations in the pulse amplitudes.

\section{Conclusion and Outlook}

To conclude, we have shown that a sequence of one-axis twisting and rotations can overcome the precision scaling of a single application of a one-axis twisting Hamiltonian, and reach the fundamental Heisenberg scaling. The approach uses the same elements as existing OAT implementations. Thus, it does not require significant modification to an existing experimental setup.

Fundamentally, the extreme spin squeezed states generated by our pulse scheme maximize metrological gain for Ramsey interferometry at a fixed signal output amplitude. They always attain Heisenberg scaling up to a constant factor, giving us additional flexibility, e.g., to limit the number of OAT pulses. This feature is useful when employing light-mediated interactions in an optical cavity to create an effective OAT as the photon scattering reduces contrast.
To account for the contrast reduction due to photon scattering, we specifically target extreme SSSs that yield a high contrast signal, requiring a reduced shearing strength $\tilde Q$. However, extreme SSSs that yield a low contrast signal could be useful for other protocols different from Ramsey scheme as they are still highly entangled, for instance, the SATIN protocol~\cite{DavisPRL2016}.

We have found that a single non-linear interaction, such as OAT, already provides a remarkable degree of control when combined with suitable non-entangling operations. This flexibility could be related to the results in Ref.~\cite{LloydPRL1995}, which shows that alternating applications of different Hamiltonians are often capable of approximating any quantum gate, or at least a relevant subset, as required here. We expect that similar strategies could be used to create other entangled states that optimize other figures of merit, such as the Fisher information or planar squeezing~\cite{vitagliano2018entanglement, birrittella2021optimal}. The latter corresponds to a modification of the squeezing parameter that quantifies the squeezed states' fitness over an extensive range of phases acquired during free evolution.

\begin{acknowledgments}
This research was supported by DEVCOM Army Research Laboratory under Cooperative Agreement Numbers W911NF-21-2-0037 (SC) and W911NF-17-2-0147 (MG).
VSM is grateful for support by a Laboratory University Collaboration Initiative (LUCI) grant from OUSD.
\end{acknowledgments}

\bibliographystyle{apsrev4-2-titles.bst}
\bibliography{refs}

\end{document}